\documentclass[pra, twocolumn, floatfix]{revtex4}
\usepackage{graphicx}
\usepackage{amsmath, amsfonts, amssymb, bm}
\usepackage{braket}
\usepackage{color}
\usepackage[utf8]{inputenc}
\usepackage{subfigure}

\begin{document}

\title{ Fragmentation of the $^4$He$_2$ dimer by relativistic highly charged projectiles in collisions with small kinetic energy release  }  
\author{ B. Najjari \!\!\!\!\! $^1$, S. F. Zhang  \!\!\!\!\! $^1$, 
X. Ma \!\!\!\!\! $^1$  
and A. B. Voitkiv \!\!$^2$ }
\email[]{alexander.voitkiv@tp1.uni-duesseldorf.de}
\affiliation{ 
$^1$ Institute of Modern Physics, Chinese Academy of Sciences, Lanzhou 730000, China \\  
$^2$ Institute for Theoretical Physics I, Heinrich-Heine-Universit\"at D\"usseldorf, Universit\"atsstrasse 1, 40225 D\"usseldorf, Germany } 



\date{\today}

\begin{abstract} 

We study theoretically the fragmentation of the helium dimer, $^4$He$_2$, into singly charged ions in collisions with relativistic highly-charged projectiles. We discuss the main mechanisms driving this process   
with the focus on the fragmentation caused by direct  ionization of both atomic sites of the dimer in a single collision with the projectile. This direct mechanism dominates the He$_2$ $\to $ He$^+$ + He$^+$ breakup events with the kinetic energies of the emerging ionic fragments below $4$--$5$ eV. 
We explore the energy and angular distributions of the He$^+$ ions produced in collisions with $1$ and $7$ GeV/u U$^{92+}$ projectiles and show that their shape is significantly affected by relativistic and higher-order effects in the interaction between the projectile and the dimer. We also show that the shape of the energy spectrum is quite sensitive to the binding energy of the He$_2$ dimer which can be exploited for its precise determination. The contribution of the direct mechanism to the total cross section for the He$_2$ fragmentation by $1$ and $7$ GeV/u U$^{92+}$ was calculated to be 
$3.65$ Mb and $2.4$ Mb, respectively, representing roughly half of this cross section.

\end{abstract}

\maketitle

\section{Introduction} 

The study of fast ion-atom and ion-molecule collisions has already a long history. However, collisions of fast ions with very extended atomic objects, like humongous   van der Waals dimers, are very rarely considered, even though  compared to the 'normal' case the physics of such collisions possesses new and interesting features.  
 
In particular, to our knowledge there are just few papers, \cite{He2_alpha-particle}-\cite{He2-rel-HCI-lett}, where collisions of fast ions with the He$_2$ and Li-He dimers were studied. 
In 
\cite{He2_alpha-particle}-\cite{thesis_He2_S14+} and \cite{He2-rel-HCI-lett} the fragmentation process 
He$_2$ $\to $ $2$ He$^+ + 2 \, \, e^{-}$ was explored for collisions with different projectiles and at different impact velocities while in \cite{He-Li-rel-HCI} single ionization of the Li-He dimer by ultra-fast projectiles involving energy transfer between the dimer atoms was considered. 

Both the He$_2 $ and Li-He dimers are spectacular 
pure quantum systems. For instance, in the He$_2$ dimer 
the interaction between two ground-state helium atoms 
is so weak that it supports one bound state only, which has  
an extremely small binding energy 
($ \simeq 10^{-7}$ eV, \cite{e-l}). The size of this 
two-atomic bound system is enormous: its average bond length is    
$ \approx 50$ \AA \, \cite{e-l} and the dimer extends to 
the distances of more than $ 200 $ \AA,  being  
the largest known ground-state diatomic molecule. 
The outer classical turning point in 
the ground state of the dimer  
is about $14$ \AA \, \cite{14A} 
that is almost four times smaller than 
its average bond length showing that the He$_2$ dimer is a quantum halo system 
which spends most of the time in the classically forbidden region. Because of its enormous size the inter-atomic interaction and the dimer binding energy are noticeably influenced by the Casimir-Polder retardation effect \cite{Cas-Pol}, \cite{retard}. 
 
\vspace{0.25cm}

When the He$_2$ interacts with external electromagnetic fields a multitude of various processes becomes possible. They include a fragmentation of the He$_2$ dimer into helium ions which may occur when the dimer is irradiated by an electromagnetic wave or bombarded by charged projectiles. 
Helium ions repeal each other that results in 
a Coulomb explosion of the system. The kinetic energy of the ionic fragments, which is released in this explosion, 
depends on their charges and the initial distance between them. The spectra of the kinetic energy release may contain a valuable information about the structure of the initial He$_2$ dimer as well 
as transient dimers (which may be formed if the time interval between the interaction with the external field and the start of  the Coulomb explosion is sufficiently large). Some of these fragmentation processes have been already studied. 

The fragmentation of the He$_2$ dimer 
into He$^+$ ions, caused by absorption 
of a single photon, was considered in 
\cite{He2_photon}- 
\cite{He2_photon_icd-calc}. 
In \cite{dimer-binding-exp} the process of 
He$_2$ fragmentation into He$^+$ ions,   
induced by absorption of two high-frequency photons, was employed to sample the dimer wave function  
at inter-atomic distances $R \gtrsim 10$ \AA  \, that enabled the authors to accurately measure the binding energy of the dimer.    

The process of He$_2$ fragmentation 
into singly charged ions in collisions with $150$ keV/u alpha particles and 
$11.37$ MeV/u S$^{14+}$ projectiles  
was explored in \cite{He2_alpha-particle} and 
\cite{He2_S14+}-\cite{thesis_He2_S14+}, respectively, 
where the focus was on  
the fragmentation events with kinetic energies of the He$^+$ ions greatly exceeding $ 1 $ eV. 

Very recently, the process of He$_2$ fragmentation in collisions with relativistic highly charged projectiles was considered in 
\cite{He2-rel-HCI-lett}. Unlike 
\cite{He2_alpha-particle} and \cite{He2_S14+}-\cite{thesis_He2_S14+}, in \cite{He2-rel-HCI-lett} the  fragmentation events with kinetic energies of He$^+$ ions not significantly exceeding $1$ eV were explored. In particular, it was shown in \cite{He2-rel-HCI-lett} that the lower-energy part of the fragmentation spectrum can be very strongly influenced by relativistic effects caused by the collision velocity approaching the speed of light and that the calculated energy spectrum is quite sensitive to the value of the dimer binding energy which can be exploited for measuring this energy. 

\vspace{0.25cm} 

In the present paper we continue to study the fragmentation reaction He$_2$ $\to$ 
$2$ He$^+$ $ + $ $2 $ e$^-$ in relativistic collisions with highly charged ions. Like in \cite{He2-rel-HCI-lett}, we will focus here on the fragmentation mechanism in which the projectile directly ionizes both helium atoms in a single collision. 
At relativistic collision  velocities this mechanism becomes 
very long-ranged and, therefore, is especially suited to probe 
the ground state of the He$_2$ at inter-atomic distances $R \gtrsim 10$ a.u., where it already 
completely dominates the fragmentation process.  
  
The paper is organized as follows. The next section begins with a discussion of the main fragmentation mechanisms which are present in collisions with relativistic highly charged ions. Then we consider a theoretical description of the direct fragmentation mechanism mentioned above. In section III we present our results for the fragmentation cross sections. Section IV contains main conclusions. 
 
Atomic units ($\hbar = \vert e \vert = m_e = 1$) 
are used throughout unless otherwise is stated.

\section{General consideration }

\subsection{ The main fragmentation mechanisms } 

Let the He$_2$ dimer collide with  
a projectile, which has a charge $Z_p$ and moves  
with a velocity ${\bm v}$ whose magnitude  
approaches the speed of light $c \approx 137$ a.u.  
At $v \sim c $ the parameter 
$\eta = Z_p/v$, characterizing the effective strength of the projectile field in the collision, is well below $1$, which indicates that the field of the projectile in the collision is on overall weak rather than strong. Consequently, the breakup of the He$_2$ will occur with a non-negligible probability only provided the number of 'steps' in the interaction of the projectile with the constituents of the dimer is reduced to a necessary minimum. Also,  
in relativistic collisions with very light atoms  
the processes of radiative and nonradiative electron capture from the atom by the projectile 
are already extremely weak having cross sections 
which are by several orders of magnitude smaller that atomic ionization and excitation cross sections \cite{el-cpt, eic, we-1998, He-ioniz}.

The above features 
inherent to high-energy collisions with light targets  
restrict the number of the main fragmentation mechanisms,  
which govern the breakup of the He$_2$ dimer into He$^+$ ions  
\begin{eqnarray}
Z_p + \text{He}_2 \to Z_p + \text{He}^+ + \text{He}^+ + 2 e^-     
\label{fm1-1}
\end{eqnarray} 
by high-energy charged projectiles, 
to four \cite{He2-rel-HCI-lett}.   

\subsubsection{ Mechanisms of the 'delayed'  fragmentation }

Two of these mechanisms can be termed as 'delayed'  
since the average time, which they require for the formation of the  He$^+$--He$^+$ system, is determined by the motion of the helium nuclei and, therefore,  exceeds by several orders of magnitude the time, which is needed by the projectile to traverse the region of space occupied by the initial He$_2$ dimer. These mechanisms are also 
characterized by relatively large values of the kinetic energy of the He$^+$ fragments. Both of them 
involve the interaction between the projectile and just one site of the He$_2$ dimer.

a) In the first of these mechanisms the collision between the projectile and one of the helium atoms results in its double ionization. Then the He$^{2+}$--He system evolves until the He$^{2+}$ radiatively captures 
one electron from the neutral helium atom and the resulting 
He$^+$-- He$^+$ system undergoes a Coulomb explosion. This mechanism -- according to its steps -- can be denoted as double-ionization--radiative-electron-transfer (DI-RET). 

The fragmentation cross section 
$ \sigma_{\text{fr}}^{\text{DI-RET}} $ for this mechanism can be evaluated as 
$ \sigma_{\text{fr}}^{\text{DI-RET}} = 2 \, \sigma_{\text{He}}^{2+} \, P_{\text{RET}}$, where  
$\sigma_{\text{He}}^{2+}$ is the cross section for double ionization of the helium atom by the projectile and $ P_{\text{RET}} $ the probability for radiative electron transfer. In relativistic collisions with very highly charged projectiles ($Z_p \sim v$) the cross section $\sigma_{\text{He}}^{2+}$ reaches several tens of megabarns which in turn results  
in the magnitude of the $ \sigma_{\text{fr}}^{\text{DI-RET}} $ of the order of megabarn (see subsection D of section III). 
   
Within the reflection approximation,   
the kinetic energy $ E_K $ of the ionic fragments, 
which is released in 
a Coulomb explosion, is related to the inter-nuclear distance 
$ R $ at which the explosion started: in particular, for singly charged ions one has $ E_K = 1/R $. 
The experimental results of \cite{dimer-binding-exp} and the theoretical consideration of \cite{we-21-reflect-appr} show that in the case of He$_2$ the reflection approximation is quite accurate up to very low energies ($E_K \sim 1$ meV) where it starts to be violated by the recoil effects.    

Since radiative electron transfer occurs only at small inter-nuclear distances (not exceeding a few atomic units) the DI-RET mechanism leads to fragmentation with relatively large kinetic energy release: the results of \cite{He2_alpha-particle} and 
\cite{thesis_He2_S14+} show that in collisions with fast charged projectiles the energy spectrum of the fragments, which are produced via this mechanism, is peaked at 
$E_K \approx 9$ eV corresponding 
to $R \approx 3$ a.u.,  and that its intensity  
rapidly decreases when the energy $E_K$ departures from 
this value. 

\vspace{0.25cm}

b) In the second 'delayed' fragmentation mechanism 
the projectile also interacts with just one helium atom. Now, however, the interaction  results in simultaneous ionization-excitation of this atom. In the presence of a neutral helium atom the excited helium ion can de-excite not only 
via spontaneous radiative decay but also by 
transferring energy  
to the neutral He atom that leads to ionization 
of the latter.  

Such an energy transfer acquires a relatively large effective range becoming  quite efficient if the He$^+$ ion is produced in an excited state which can decay to the ground state by an electric dipole transition. Then the energy transfer occurs mainly via dipole-dipole two-center electron transitions. 

To our knowledge, a radiationless relaxation of an excited atom via energy transfer to a neighbor atom, which at large inter-atomic distances occurs predominantly via the dipole-dipole interaction, was first considered theoretically in 
\cite{i-a-a}. In the process, which was studied in \cite{i-a-a}  and termed there {\it inter-atomic Auger decay} (IAAD),   
the exited atom has initially a vacancy which is filled by its electron,   
but -- unlike in 'normal' (intra-atomic) Auger decay --  the energy release is transferred to the neighbor atom ionizing it. 
The process of IAAD was shown in \cite{i-a-a} to be so  
efficient that it can even outperform intra-atomic Auger decay. 

In \cite{i-a-a} also  
a simple approximate formula was derived for the IAAD rate $ \Gamma_a $. Being adapted to the case under consideration, the formula reads    
\begin{eqnarray}
\Gamma_a = \frac{ 3 \alpha }{ 8 \pi } \, \,  
\left( \frac{ c }{ \omega } \right)^4 \, \, \,   
\frac{ \Gamma_r \, 
\sigma_{\text{ph}}^{\text{He}}(\omega) }{ \, R^6}.
\label{two-center-rates}
\end{eqnarray} 
Here, $ \Gamma_r $ is the rate for spontaneous radiative decay of the excited state of the He$^+$ ion, 
$\omega$ is the energy difference between the excited and ground states of the He$^+$ ion which is transferred to the He atom, 
$ \sigma_{\text{ph}}^{\text{He}}(\omega) $ is the cross section for single ionization of the He atom by absorption of a photon with frequency $\omega$, $ R $ 
is the distance between the He$^+$ and He, and $\alpha$ is a numerical parameter $\sim 1 $ depending on the magnetic quantum number of the excited state of the He$^+$.    

The process of IAAD in a system of two atomic particles, where relaxation via intra-atomic Auger decay is not allowed energetically, was computed in \cite{icd} and was 
called there {\it interatomic coulombic decay} (ICD).    
Nowadays the term ICD is often used 
to denote relaxation mechanisms, where  
an overlap between electronic shells of the interacting particles is not required since the energy is transferred via the exchange of virtual photons between them. 
Accordingly, the fragmentation mechanism, in which simultaneous ionization-excitation of one of the atoms by the projectiles is followed by ICD decay, can be denoted as IE-ICD. 

The process of ICD in the He$^+$--He system 
competes with spontaneous radiative decay, which does not result in the production of the second helium ion. Using Eq. (\ref{two-center-rates}) and assuming for definiteness that the excited He$^+$ ion was produced in a $2p$--state we obtain for the ratio 
$\Gamma_a/\Gamma_r \approx \beta \, (10/R)^6 $, where $\beta \sim 1$. This indicates that when the distance $R$ between the He$^+$ and He significantly exceeds $10$ a.u. the ICD channel becomes inefficient (in the experiment 
\cite{He2_photon_icd} on the He$_2$ photo fragmentation the ICD was clearly 'visible' up to $R \approx 12$ a.u.). According to the results of 
\cite{He2_alpha-particle} and \cite{thesis_He2_S14+}  in fast collisions with charged projectiles the IE-ICD is very efficient in producing He$^+$ ions with energies in the vicinity of $E_K \approx 8$ eV, but is in essence completely inactivated above $10$ eV and 
below $ 4$--$5$ eV. As rough estimates suggest  
(see subsection D of section III), in relativistic collisions with very highly charged projectile the sum of the contributions $ \sigma_{\text{fr}}^{\text{IE-ICD}}$ and 
$\sigma_{\text{fr}}^{\text{DI-RET}}$
to the total fragmentation cross section may rich few megabarns.    

\vspace{0.25cm} 

c) The average size of the He$_2$ dimer greatly exceeds not only 
the effective range of RET but also that of ICD. 
Therefore, there must occur a very significant contraction of the intermediate (He$^+$)$^*$--He and He$^{2+}$--He dimers 
before the distance between the nuclei 
sufficiently decreases in order for 
the ICD and RET to come into the play. 
The contraction is possible because the potential of the He$^{2+}$--He and (He$^{+}$)$^*$--He systems at large inter-nuclear distances is attractive.

\subsubsection{ Mechanisms of the 'instantaneous'   fragmentation  }

In collisions with high-energy projectiles there are two more fragmentation mechanisms in which, unlike the DI-RET and IE-ICD, the He$^+$-He$^+$ system emerges  'instantaneously', i.e. on the so short time scale that during the production of two helium ions the helium nuclei remain essentially at rest.     

   a). In one of them the projectile 
interacts with both atoms of the dimer. As a result, each helium atom emits an electron and becomes  
a singly charged ion. In this fragmentation mechanism the projectile directly 
forms the He$^+$ - He$^+$ system 
\cite{Sulf} which undergoes a Coulomb explosion. Since the size of the He$_2$ is very large, the interactions between the constituents of 
the dimer play in this mechanism no noticeable role.  
In \cite{He2-rel-HCI-lett} this mechanism was termed 
{\it the direct fragmentation} (DF) 
and we shall use it also here. 

  We note that at relativistic impact velocities the time interval $T$ between the collisions of the projectile with the first and second atoms of the He$_2$ dimer 
does not exceed few atomic units ($T \lesssim 10^{-16}$ s). 
This is by orders of magnitude smaller than 
typical nuclear times ($\sim 10^{-13}$--$10^{-12}$ s)
in the He$^+$ -- He system,   
showing that the DF mechanism is indeed 'instantaneous'. 

\vspace{0.25cm} 

  b). In the other 'instantaneous'  fragmentation mechanism the projectile interacts with just one atom 
of the dimer. As a result of this interaction,  
the atom emits an electron. There is a certain probability that the emitted electron will move towards the other atom and 
knock out one of its electrons. 
Thus, this fragmentation mechanism is a combination of single ionization 
of the helium atom by a high-energy projectile 
and the so called e-2e process on helium 
(single ionization by electron impact). 
Following \cite{He2-rel-HCI-lett} we shall refer to this mechanism as   
{\it single ionization -- e-2e } (SI--e-2e) \cite{f1}. 

In the SI--e-2e process the velocity of the first emitted electron is by two orders of magnitude smaller than the velocity of the relativistic projectile. Nevertheless, this electron still moves much faster than 
the helium nuclei. With its typical velocity $v_e \sim 1$ a.u., we obtain that the time, which the electron needs to propagate between the sites of the dimer, does not exceed $ 10^{-14} $ s. 
Since this time is much shorter than typical nuclear times 
in the He$^+$ -- He system, the SI--e-2e mechanism can be also viewed as 'instantaneous'.  

\vspace{0.25cm} 

  Since in both the DF and SI--e-2e 
the helium ions emerge at 'frozen' positions of the helium nuclei, their energy $E_K$ is very simply ($E_K = 1/R $) related to the size $R$ of the He$_2$ dimer at 
the collision instant (as long as the recoil effects can be ignored, see subsection B of the present section). 

  The cross section for the production of two singly charged helium ions by relativistic projectiles via the DF mechanism depends on the transverse size $R_{\perp}$ of the He$_2$ dimer \cite{He2-rel-HCI-lett}, scaling approximately as $ 1/R_{\perp}^2 $. The corresponding cross section for the production via the SI--e-2e depends on the dimer size $R$ as $1/R^2$ \cite{He2-rel-HCI-lett}.  

  Thus, unlike the DI-RET and IE-ICD, the 
DF and SI--e-2e mechanisms are not only instantaneous, providing a simple correspondence between the kinetic energy release and the size of the He$_2$ dimer, but also possess much longer effective range. Therefore, they are especially suited for probing the ground state of the He$_2$ in a very large interval of the inter-atomic distances $R$. 


Our calculations show that in collisions of He$_2$ dimers with very highly charged projectiles ($Z_p/v \sim 1 $) the DF mechanism is much more efficient than the SI--e-2e yielding cross sections which are larger by two orders of magnitude. Therefore, in the rest of the paper we shall concentrate on the He$_2$ fragmentation via the DF mechanism.

\subsection{ The direct fragmentation mechanism }
   
We shall consider collisions between the He$_2$ dimer 
and the projectile using the semi-classical 
approximation in which the relative motion of the heavy particles (nuclei) is treated classically whereas the electrons are considered quantum mechanically. 
  
We choose a reference frame in which the dimer is at rest and take the nucleus of one of its atoms as the origin. We shall refer to this atom as atom $A$ whereas the other will be denoted by $B$. In this frame, the coordinates of the nucleus of atom $B$ are given by the inter-nuclear vector 
${\bm R}$ of the dimer and the projectile moves along a classical straight-line trajectory $ \mathbf{ R }_p (t) = \bm b + \bm v t$, where 
$\bm b = (b_x, b_y, 0)$ is the impact parameter 
with respect to the nucleus of atom $A$ and $\bm v = (0,0,v)$ the collision velocity.   

Since the size of the He$_2$ dimer is very large, the ionization of atoms $A$ and $B$ occurs independently of each other and we can use the independent electron approximation. According to it the probability $ P_{A^+B^+}$ for single ionization of atoms $A$ and $B$ in the collision where the projectile moves with an impact parameter $\bm b$ is given by  
\begin{eqnarray} 
P_{A^+B^+} =  P_{A^+}(\bm b) P_{B^+}(\bm b').  
\label{indep-1} 
\end{eqnarray} 
Here, $P_{A^+}(\bm b)$ and $P_{B^+}(\bm b')$ are the probabilities for single ionization of atoms $A$ and $B$, respectively, and 
${\bm b'} = {\bm b} - {\bm R}_{\perp}$ is the collision impact parameter with respect to atom $B$, where 
${\bm R}_{\perp}$ is the part of the inter-nuclear 
vector $ {\bm R} $ of the dimer which is perpendicular to the projectile velocity ${\bm v}$.    
  
Within the independent electron approximation 
the probabilities for single ionization of helium atoms $A$ and $B$ read 
\begin{eqnarray} 
P_{A^+}(\bm b) & = & 2 \, w(\bm b) \, (1 - w(\bm b)) 
\nonumber \\ 
P_{B^+}(\bm b') & = & 2 \, w(\bm b') \, (1 - w(\bm b')),  
\label{indep-2} 
\end{eqnarray}
where $ w(\bm b) $ ($ w(\bm b') $) is the probability to remove one electron from helium atom in the collision with a projectile having an impact parameter $\bm b$ ($\bm b'$) \cite{new-approach}.  

The quantity 
\begin{eqnarray} 
\sigma^{\text{DF}}_{A^+B^+} & = & \int d^2 {\bm b} \, P_{A^+B^+}(\bm b) 
\nonumber \\ 
   & = & \int d^2 {\bm b} \, P_{A^+}(\bm b) \, 
P_{B^+}(\bm b')  
\nonumber \\ 
& = & \int d^2 {\bm b} \, P_{A^+}(\bm b) \, 
P_{B^+}({\bm b} - {\bm R}_{\perp})
\label{indep-3} 
\end{eqnarray}
represents the cross section for the production of two singly charged helium atom by the projectile in the collision with the He$_2$ dimer at a given inter-nuclear vector ${\bm R}$ of the latter. Since the probabilities 
$ P_{A^+}(\bm b) $ and $ P_{B^+}(\bm b') $ depend just on the absolute value of the respective impact parameters, 
$ P_{A^+}(\bm b) \equiv P_{A^+}(b)$ and 
$ P_{B^+}(\bm b') \equiv P_{B^+}(b') $, 
the cross section (\ref{indep-3}) depends only 
on the absolute value $ R_{\perp} $  of the two-dimensional vector $ {\bm R}_{\perp}$:  
$\sigma^{\text{DF}}_{A^+B^+} = \sigma^{\text{DF}}_{A^+B^+}(R_{\perp}) $.   
  
In our calculations of the probabilities $w(b)$ and $w(b')$ 
we regarded the helium atom in its initial and final states 
as an effectively single-electron system,   
where the 'active' electron moves in the effective 
field, created by the 'frozen' atomic core consisting 
of the atomic nucleus and the 'passive' electron. 
This field is approximated by the potential  
\begin{eqnarray} 
V( {\bm r} ) = - \frac{ 1 }{ r } - \left(1 + \beta r \right) %
\frac{ \exp(- \alpha r) }{ r }, 
\nonumber       
\label{potential} 
\end{eqnarray}   
where $r$ 
is the distance between the active electron 
and the atomic nucleus, and 
$\alpha=3.36$ and $\beta=1.665$ \cite{param}. 
We note that with this choice of $\alpha$ and $\beta$, 
$V({\bm r})$ almost coincides with 
the exact Hartree-Fock potential. 

The cross section (\ref{indep-3}) can be only computed numerically and the computation is quite time consuming. However, a simple estimate can be obtained for this cross section at $R_\perp \gg 1$ a.u. \cite{He2-rel-HCI-lett} 
\begin{eqnarray} 
\sigma^{\text{DF}} \approx C \, \frac{ Z_p^4 }{ v^4 \gamma^2 } \, \, \bigg[ K^2_1\left( \frac{ R_\perp }{ R_a } \right) + \frac{ 1 }{ \gamma^2 } \, 
K^2_0\left( \frac{ R_\perp }{ R_a } \right) \bigg].  
\label{sech-analytic} 
\end{eqnarray} 
Here, $ K_0 $ and $ K_1 $ are the modified Bessel function \cite{A-S}, 
$ R_a  = \frac{ \gamma v }{ \overline{ \omega } } $ is the adiabatic collision radius, where  $ \gamma = 1/\sqrt{1-v^2/c^2} $ is the collision Lorentz factor,   
$\overline{ \omega } \approx 1.2 $ a.u. is the mean transition frequency for single ionization of a helium atom, 
and $C$ is a free 
(fit) parameter which is a very slowly varying function of $R_\perp$ \cite{param-C}.    
   
Taking into account that $K_0(x) \sim \ln (1.12/x) $ and $K_1(x) \sim 1/x$ if $ x < 1 $ whereas $K_0(x) \sim K_1(x) \sim \sqrt{ \frac{\pi}{2 x }} \exp(-x)$ at $ x > 1$, it follows from Eq. (\ref{sech-analytic}) that at $ 1 \ll R_{\perp} \lesssim R_a $ the cross section 
decreases with $R_\perp$ relatively slowly, 
$ \sigma^{\text{DF}} \sim R^{-2}_{\perp} $, whereas 
at $ R_\perp > R_a $ the decrease is already exponential. 
This means that the projectile is able to 
efficiently irradiate both atoms of the dimer 
only provided that the dimer transverse size $R_\perp$ 
does not exceed the adiabatic collision radius $R_a$. 
Since $R_a \sim \gamma v$,  
an ultrafast projectile possesses a very large effective interaction range and can probe systems with large dimensions.     
Simple estimates show that, beginning with impact energies 
of a few GeV/u, the adiabatic collision radius exceeds the size of the He$_2$ dimer \cite{He2-rel-HCI-lett}.  

\vspace{0.25cm} 

The cross section (\ref{indep-3}) refers to the situation where two helium atoms in the dimer are separated by the vector ${\bm R}$ with the absolute value of its transverse projection $R_{\perp}$. It yields an important information about the collisions giving an insight into the basic physics of the fragmentation process \cite{He2-rel-HCI-lett} but cannot be  
(directly) measured in experiment.  
Therefore, let us 'convert' the cross section (\ref{indep-3}) into quantities which can be measured. We start with the expression   
\begin{eqnarray}      
\frac{ d \sigma^{\text{DF}}_{\text{fr}}}{ d^3 {\bm R} } = 
\sigma^{\text{DF}}_{A^+B^+}(R_{\perp}) \, \, \big| \Psi_i({\bm R}) \big|^2,   
\label{convert-1} 
\end{eqnarray}
where $\Psi_i({\bm R}) $ is the wave function describing 
the relative motion of the atoms in the ground state of the He$_2$ dimer.  

\vspace{0.25cm} 

  As was already mentioned, we assume that before the collision with the projectile the He$_2$ dimer was at rest. 
Let ${\bm P}^{\text{rec}}_A$ and 
${\bm P}^{\text{rec}}_B$ be the recoil momenta of the helium ions, which they acquire during the collision because of the interaction with the projectile and electron emission, and let ${\bm P}_A$ and ${\bm P}_B$ be the final momenta of these ions after the Coulomb explosion. Then the energy conservation for the relative motion of the ionic fragments reads  
\begin{eqnarray} 
E_{K}  = \frac{ \big[ \frac{ 1 }{ 2 } 
\left( {\bm P}^{\text{rec}}_A - {\bm P}^{\text{rec}}_B\right) \big]^2  }{ 2 \mu } +  
\frac{ Q_A Q_B }{ R }.       
\label{convert-2} 
\end{eqnarray} 
Here, $E_K = {\bm K}^2/(2 \mu)$ is the final kinetic energy of the relative motion of the ionic fragments,  
${\bm K} = \left( {\bm P}_A - {\bm P}_B \right)/2 $ and 
$ \mu $ are the momentum of their relative motion and the reduced mass, respectively,    
$Q_A $ and $ Q_B$ ($Q_A = Q_B = 1$) are their charges and $R$ is the distance between the ions when the Coulomb explosion began and which in the case under consideration coincides with the size of the initial He$_2$ dimer at the collision instant.  In the energy balance (\ref{convert-2}) we have neglected the kinetic energy of the He$^+$ ions, which they had due to the nuclear motion before the collision and which is very small because the depth of the potential well in the He$_2$ dimer is just $\approx 1$ meV.        
 
Since in the process of single ionization of helium atoms by relativistic projectiles 
the recoil momenta of the helium ions do not noticeably exceed $1$ a.u. \cite{rhci}-\cite{r-sea},   
the first term on the right hand side of 
Eq. (\ref{convert-2}) is in the meV range. Such values are comparable to the Coulomb potential energy $Q_A Q_B/R$ at the inter-nuclear distances $R$ of the order of $10^4$ a.u. which is much larger than the size of the He$_2$ (and any known) dimer. Therefore, the neglect of the recoil energy, given by the first term on the right-hand side of Eq. (\ref{convert-2}), does not have any substantial 
impact on the energy balance. Thus, the relation 
\begin{eqnarray} 
E_{K}  =  \frac{ Q_A Q_B }{ R }       
\label{convert-3} 
\end{eqnarray}  
between the kinetic energy release and the size of the dimer at the collision instant, which neglects the recoil energies,    
is expected to be very accurate down to energies 
$E_K \simeq 10$ meV. 

\vspace{0.25cm} 

Let us now make a more restrictive assumption 
that not only the recoil energies are much smaller than the Coulomb energy $ Q_A Q_B/R $ but also that the absolute values of the recoil momenta,  $\vert {\bm P}^{\text{rec}}_A\vert $ and $\vert {\bm P}^{\text{rec}}_B \vert$, of the fragments are significantly less than $K = \vert {\bm K} \vert $. Since these values do not exceed $1$ a.u. the assumption will be fulfilled beginning with $K \gtrsim 4$--$5$ a.u. that corresponds to the energies 
$E_K \gtrsim  60$ meV.  
Under such conditions   
the momentum $\bm K$ will be directed 
essentially along the inter-nuclear vector $\bm R$ of the initial He$_2$ dimer and, taking also into account (\ref{convert-3}), we get  
\begin{eqnarray} 
d^3 \bm R &=& 
\frac{ (Q_A Q_B)^3 }{ \mu \, K \, E^4_{K} } 
\, \, d^3 {\bm K}.         
\label{convert-4} 
\end{eqnarray}   
Using (\ref{convert-1}) and (\ref{convert-3})-(\ref{convert-4}) 
we obtain 
the fragmentation cross section differential in the relative momentum ${\bm K} $ of the ionic fragments  
\begin{eqnarray} 
\frac{ d \sigma^{\text{DF}}_{\text{fr}}}{ d^3 {\bm K} } 
  & = & \frac{ (Q_A Q_B)^3  }{ \mu \, K \, E^4_{K} } \, 
\, \, \sigma^{\text{DF}}_{A^+B^+} 
\big( R_{\perp}(\bm K)\big) \, 
\nonumber \\ 
& & \times \bigg| \Psi_i\bigg( \frac{ Q_A Q_B}{E_K} \hat{\bm K}  \bigg) \bigg|^2.    
\label{convert-5} 
\end{eqnarray}
where $\hat{\bm K} = {\bm K}/K$,  
$ R_{\perp}(\bm K) = (Q_A Q_B \, \sin \Theta_{\bm K})/ E_K   $ and $\Theta_{\bm K}$ is the polar angle of the momentum $\bm K$ (the $z$-axis is along the projectile velocity $ \bm v$). 
Since the bound state of the He$_2$ dimer is spherically symmetric, the wave function 
$\Psi_i $ does not depend on $\hat{\bm K}$: 
$\Psi_i\bigg( \frac{ Q_A Q_B}{E_K} \hat{\bm K}  \bigg) = 
\Psi_i\bigg( \frac{ Q_A Q_B}{E_K} \bigg)$.   
 
Taking into account that $ d^3 {\bm K} = K^2 dK \, d \Omega_{\bm K} = \mu K d E_K \, \sin \Theta_{\bm K} d{\bm K} d \varphi_{\bm K} $, where 
$ \varphi_{\bm K}$ is the azimuthal angle of $\bm K$, and that the right-hand side of 
Eq. (\ref{convert-5}) does not depend on $ \varphi_{\bm K}$ we obtain the fragmentation cross section differential in the kinetic energy release $E_K$ and the polar angle $\Theta_{\bm K}$  
\begin{eqnarray} 
\frac{ d \sigma^{\text{DF}}_{\text{fr}}}{ dE_K d \Theta_{\bm K} } 
& = & 2 \pi \, \frac{ (Q_A Q_B)^3  }{ E^4_{K} } \, 
\bigg| \Psi_i\bigg( \frac{ Q_A Q_B}{E_K} \bigg) \bigg|^2   
\nonumber \\ 
& \times & \sin \Theta_{\bm K} 
\, \, \sigma^{\text{DF}}_{A^+B^+} 
\big( Q_A Q_B \sin \Theta_{\bm K}/E_K\big).    
\label{convert-6}  
\end{eqnarray} 

In our consideration of the direct fragmentation process, its two steps -- 'instantaneous' production of two He$^+$ ions and the consequent Coulomb explosion of the He$^+$--He$^+$ system -- are fully disentangled, even though the energy of the emitted electrons and the kinetic energy release have the same source: the energy transferred by the projectile to the He$_2$ dimer. However, since the electrons emitted from helium atoms in collisions with relativistic projectiles have energies extending up to a few tens of eV whereas, as will be seen below, the spectrum of the kinetic energy release in the DF process peaks at $E_K < 1$ eV and essentially vanishes already at $E_K  \simeq 5$--$6$ eV, such a consideration is expected to be quite accurate.

\section{ Results and Discussion }   

\subsection{Preliminary remarks} 
  
In this section we report our results for the fragmentation cross sections. In  our calculations the ground state of the He$_2$ dimer was described by a wave function $ \Psi_i $ corresponding to the dimer binding energy $I_b$ 
of $ 139 $ neV (this value is very close to that given in \cite{139.2}): this wave function was used to get results presented below in figures 1-7. In addition, in order to explore the sensitivity of the cross sections to the value of $I_b$, also wave functions $ \Psi_i $ corresponding to $I_b = 130$ neV and $148$ neV were applied for obtaining results shown in figure \ref{figure8}. 

Four different approximations were employed for computing the probabilities $w $ for single electron removal from a helium atom (see Eq. (\ref{indep-2})). 
They include: 
i) the nonrelativistic first Born approximation (nr-FBA);  ii) the nonrelativistic symmetric eikonal  approximation (nr-SEA) \cite{nr-sea}; 
iii) the relativistic first Born approximation (r-FBA) and  iv) the relativistic symmetric eikonal  approximation (r-SEA) \cite{r-sea}.  
 
The differences between results obtained by using the first Born and symmetric eikonal approximations  offer an idea about the importance of the higher-order effects in the interaction between the projectile and the atoms of the dimer. On the other hand, the differences between results of the relativistic and nonrelativistic approximations yield the information about the role of relativistic effects in the He$_2$ fragmentation.   

In particular, in the nr-FBA and nr-SEA the speed of light $c$ is assumed to be infinite which means that 
all relativistic effects caused by very high collision velocities vanish. 
In the r-FBA and r-SEA $c \approx 137$ a.u. and relativistic effects arising due to high impact energies are taken into account. Since even in very high-energy collisions with helium targets the overwhelming majority of emitted electrons have energies well below $100$ eV 
\cite{we-jpb-2005},  the main relativistic effects, which are due to high impact energies and which may influence the He$_2$ fragmentation, are caused by the deviation of the electric field of the projectile from the (unretarded) Coulomb form.  

One should say that,  
in addition to relativistic effects due to high impact velocities, there are also relativistic effects in the ground state of the free He$_2$ which -- according to \cite{139.2} --  reduce the binding energy of the dimer by about $14 \% $. This affects the form of its wave function (especially at very large inter-atomic distances) that in turn may have an impact on the shape of the spectrum of the fragments \cite{He2-rel-HCI-lett}.  

\subsection{ Fragmentation spectra }  

\begin{figure}[h!]
\vspace{0.2cm} 
\hspace{-0.5cm}
\includegraphics[width=9.cm]{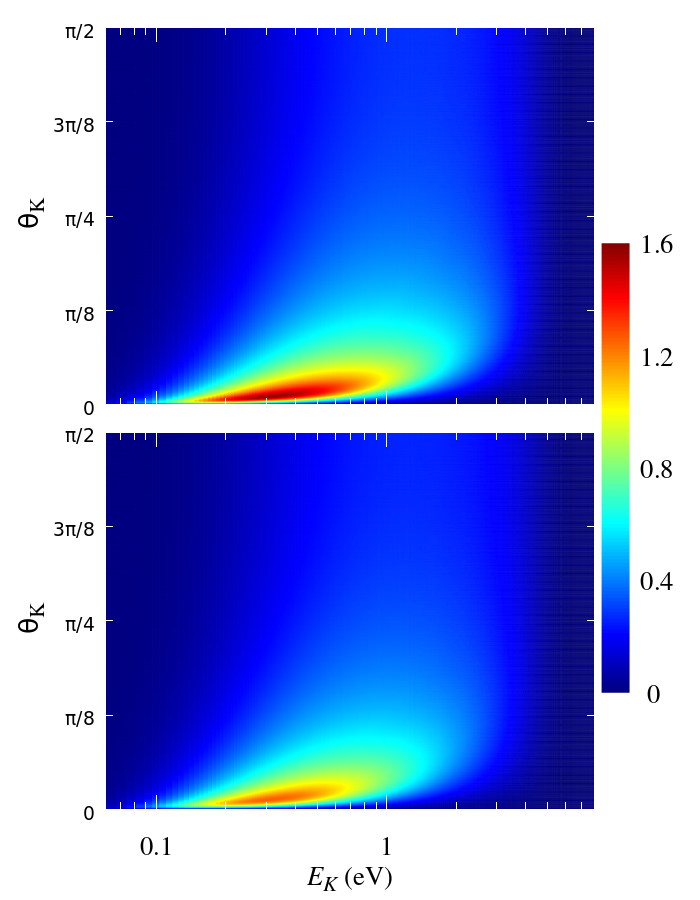} 
\vspace{-0.55cm}
\caption{ The cross section 
$ \frac{d \sigma^{\text{DF}}_{\text{fr}}}{ dE_K d \Theta_{\bm K} } $ for the fragmentation of the He$_2$ dimer by $1$ GeV/u U$^{92+}$ projectiles 
given as a function of the kinetic energy release 
$E_K$ and the angle $ \Theta_{\bm K}$. The upper and lower panels shown results obtained using the r-FBA and r-SEA, respectively. }
\label{figure1}
\end{figure}
\begin{figure}[h!]
\vspace{-0.15cm}
\hspace{-0.5cm}
\includegraphics[width=9.cm]{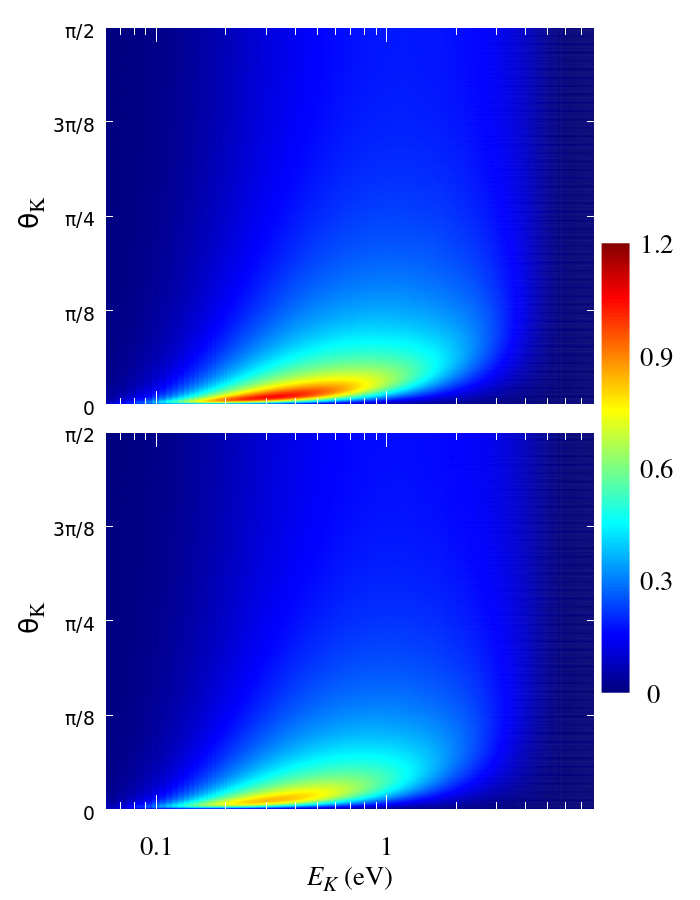} 
\vspace{-0.55cm} 
\caption{ Same as in figure \ref{figure1} but 
for the fragmentation by $7$ GeV/u U$^{92+}$. }
\label{figure2}
\end{figure}

In figures \ref{figure1} and \ref{figure2} 
we show the doubly differential cross section 
$ \frac{d \sigma^{\text{DF}}_{\text{fr}}}{ dE_K d \Theta_{\bm K} } $ 
for the fragmentation of the He$_2$ dimer into 
He$^+$ ions occurring in collisions with $1$ GeV/u and $7$ GeV/u U$^{92+}$ projectiles, respectively. The corresponding collision velocities and the Lorentz factor 
are $v = 120 $ a.u., $\gamma = 2.08$ and 
$v = 136$ a.u., $ \gamma = 8.5$, respectively. 
The cross section is plotted as a function of the kinetic energy release $E_K$ and the fragmentation angle $\Theta_{\bm K}$ presenting a general picture of the fragmentation spectrum. Taking into account that the spectrum is symmetric with respect to the transformation 
$ \Theta_{\bm K} \leftrightarrow \pi - \Theta_{\bm K} $, only the interval of angles $ 0 \leq \Theta_{\bm K} \leq \pi/2 $ is considered.  

It is seen in these figures that the spectrum is predominantly localized in the energy interval $ 0.1 $ eV $\lesssim E_K \lesssim 2$ eV and 
at fragmentation angles $\Theta_{\bm K}$ not significantly exceeding $20^\circ$. It is also seen that at very small energies $E_K$ the spectrum is restricted to smaller fragmentation angles 
$\Theta_{\bm K}$ whereas with increasing the energy the spectrum shifts to larger 
$\Theta_{\bm K}$. 

The interval of kinetic energies 
$ 0.1 $ eV $\lesssim E_K \lesssim 2$ eV corresponds 
to the instantaneous size $R$ of the He$_2$ dimer during the collision in the range  
$ 14 $ a.u. $\lesssim R \lesssim 270 $ a.u. 
The probability to find the He$_2$ dimer 
in this range is above $90 \%$ that is reflected in the dominance of the above energy interval in the spectrum. At energies $ \gtrsim 5$--$6$ eV the spectrum practically vanishes since they correspond to the inter-nuclear distances $ R \lesssim 4$ a.u.,  where the probability to find the He$_2$ dimer is negligibly small.  
 
The cross section $ \sigma^{\text{DF}}_{A^+B^+} $ 
for the production of two helium ions by the projectile decreases with the transverse size $R_{\perp}$ of the dimer at the collision instant 
($ \sigma^{\text{DF}}_{A^+B^+} \sim 1/R_{\perp}^2$ at $R_{\perp} < R_a$).   
Since at a fixed 
$\Theta_{\bf K}$ the value of   
$R_{\perp}$ is proportional to $R$, the fragmentation with very small $E_K$ may occur at very small   
$\Theta_{\bf K}$ only, whereas at significantly larger values of $E_K$ it becomes possible  
at not very small $\Theta_{\bf K}$ as well. 
Taking also into account that the cross section 
$ \frac{d \sigma^{\text{DF}}_{\text{fr}}}{ dE_K d \Theta_{\bm K} } $ is proportional to the geometrical factor $\sin \Theta_{\bm K}$ we can explain the 'shift' of the spectrum to larger 
$\Theta_{\bm K}$ with increase in $E_K$, which is observed in figures \ref{figure1} and \ref{figure2}.   
  
It also follows from the results presented in 
figures \ref{figure1} and \ref{figure2} that for collisions with so very highly charged projectiles, as U$^{92}$ ions, the symmetric eikonal approximation predicts a noticeably smaller number of the fragmentation events. This shows that the higher-order effects in the interaction between the projectile and the target remain visible even at very high impact energies. The smaller cross section values, predicted by the symmetric eikonal approximation, reflects the general tendency that in high-energy collisions the higher-order effects 
reduce cross section values.  

\vspace{0.25cm} 

\begin{figure}[h!]
\vspace{-0.25cm}
\centering
\includegraphics[width=9.2cm]{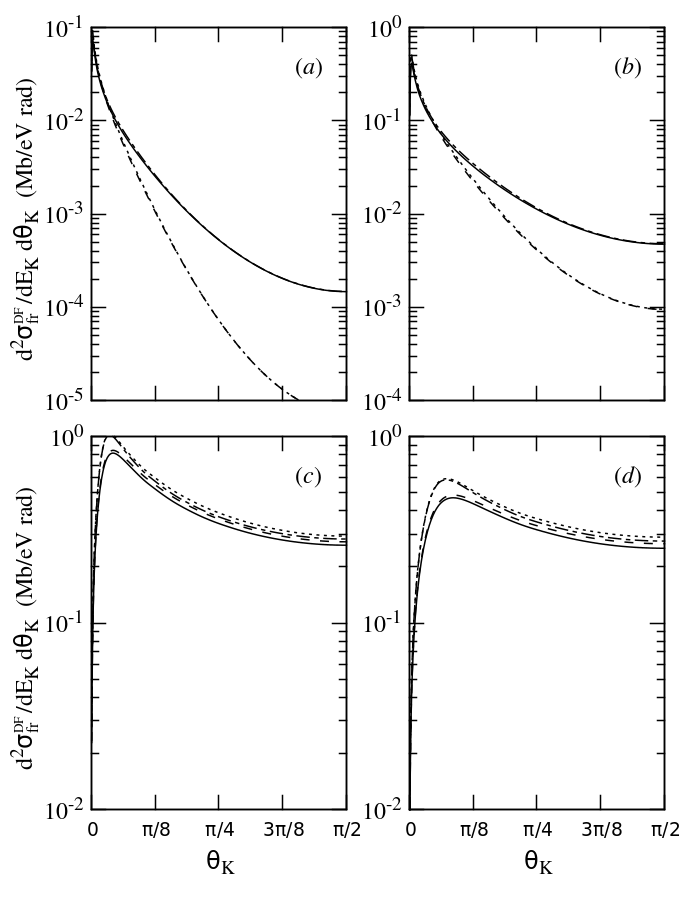} 
\vspace{-0.55cm}
\caption{ The cross section $ \frac{d \sigma^{\text{DF}}_{\text{fr}}}{ dE_K d \Theta_{\bm K} } $ for the fragmentation of the He$_2$ dimer by $1$ GeV/u U$^{92+}$ projectiles 
given as a function of the angle $ \Theta_{\bm K}$  at fixed values of the energy $E_K$: $60$ meV ($a$), $100$ meV ($b$), $1$ eV ($c$) and $2$ eV ($d$). Dotted, dashed, dash-dotted and solid curves are results of the calculations in the nr-FBA, the nr-SEA, the r-FBA and the r-SEA, respectively. }
\label{figure3}
\end{figure}

\begin{figure}[h!]
\centering 
\vspace{-0.25cm}
\subfigure{\includegraphics[width=9.cm]{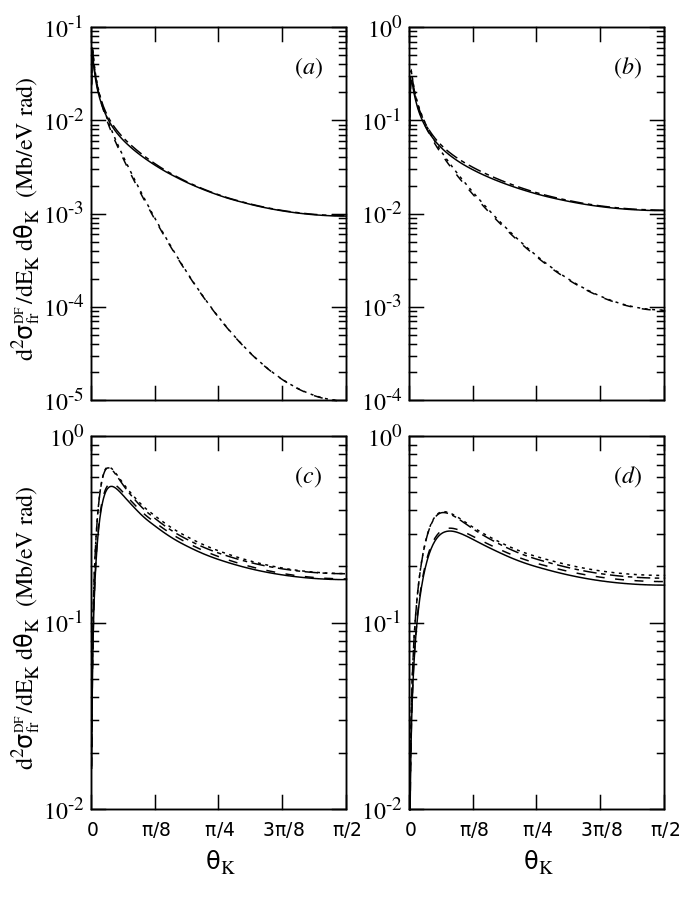}} 
\vspace{-0.55cm}
\caption{ Same as in figure \ref{figure3} but for collisions with $7$ GeV/u U$^{92+}$ projectiles.  }
\label{figure4}
\end{figure}

Figures \ref{figure3} and \ref{figure4} give a more detailed information about the fragmentation process by displaying the cross section 
$ \frac{d \sigma^{\text{DF}}_{\text{fr}}}{ dE_K d \Theta_{\bm K} } $ 
as a function of the fragmentation angle 
$\Theta_{\bm K}$ at a given value of 
the kinetic energy release $E_K$: $60$ meV, $100$ meV, $1$ eV and $2$ eV. 
Some conclusions can be drawn from the results presented in these figures.  

First, by comparing results of the different  calculations we see that the shape of the He$_2$ fragmentation 
pattern is in general influenced by both relativistic effects, which arise due to collision velocities approaching the speed of light, and higher-order effects, which originate in a very high charge of the projectile.  

Second, the influence 
of the relativistic effects increases when the kinetic energy release decreases: these effects especially impact  
the fragmentation events with very small kinetic energy release 
where they enhance the cross section up to one or two orders of magnitude. On the other hand, the fragmentation with larger values of $E_K$ is only very modestly influenced by these effects. 

Third, the situation with the higher-order effects is just opposite: their role rises with increasing 
$E_K$ and they mostly affect the fragments having larger energies. However, even for such fragments these effects  remains modest, reducing the cross section values not significantly more than $10$-$20 \%$. The higher-order effects also somewhat influence the low-energy part of the spectrum: in this case, however, their effect is limited to the very small angles $\Theta_{\bm K}$ only. 

The above qualitative features 
of the fragmentation process can be understood by noting that the cross section $ \sigma^{\text{DF}}_{A^+B^+}(R_{\perp}) $ 
is most profoundly influenced by the relativistic effects when the magnitude of $R_{\perp}$ is large 
($ R_a /\gamma < R_{\perp} \lesssim R_a $)         
whereas the higher-order effects are most significant in collisions where the projectile passes close to both atoms of the dimer (the impact parameters $b$ and $b'$ do not exceed a few atomic units)  that is possible only at 
relatively small values of $R_{\perp}$.  

Very large values of $R_{\perp}$  
imply that the instantaneous dimer size $R$ in the collision has to be very large and, in addition, the dimer orientation angle $ \Theta_{\bf R} $ should not be small. Since $E_K = 1/R$ and 
$ \Theta_{\bf R} \approx \Theta_{\bf K}$, it follows that the relativistic effects are most significant for fragmentation events, which are characterized by a very small kinetic energy release and not too small angles $\Theta_{\bm K}$. 

On the other hand, collisions occurring at small values of $R_{\perp}$, where the higher-order effects can be substantial, are characterized by either small values of the instantaneous dimer size $R$ or very small angles $\Theta_{\bm R} \approx \Theta_{\bm K}$ (or a  combination of both). The kinetic energy release 
$ E_K $ of $0.06$ eV, $0.1$ eV, $1$ eV and $2$ eV  corresponds to $R \approx 453$ a.u., $ 272 $ a.u., 
$ 27.2 $ a.u. and $ 13.6 $ a.u., respectively. In order to have substantial higher-order effects at the first two values of $R$  
the dimer has to be almost parallel to the collision velocity. 
As a result, in collisions with very small $E_K$ such effects can become visible only at the fragmentation angles $\Theta_{\bm K} $ close to zero whose contribution, due to the geometric factor $\sin \Theta_{\bm K}$, is very small. For collisions with larger $E_K$ the restrictions on the angle $\Theta_{\bm K}$ are not so strict and the higher-order effects may be noticeable for the whole range $ 0^\circ \leq \Theta_{\bm K} \leq 90^\circ $.  

The energy $E_K $ of $0.06$ eV, $0.1$ eV, $1$ eV and $2$ eV correspond to the relative momentum $K$ of $\approx 4$ a.u., $ 5.2 $ a.u., 
$ 16.4 $ a.u. and $ 23.2 $ a.u., respectively. The first two momentum values are not very large. Therefore, in the case of $E_K = 0.06$ eV and $0.1$ eV the very narrow peak at very small angles 
(see panels ($a$) and ($b$) of figures \ref{figure3}--\ref{figure4}) can be noticeably smeared out by the recoil effects.   

\vspace{0.25cm}

\begin{figure}[h!]
\vspace{0.0cm} 
\centering
\subfigure{\includegraphics[width=9.5cm]{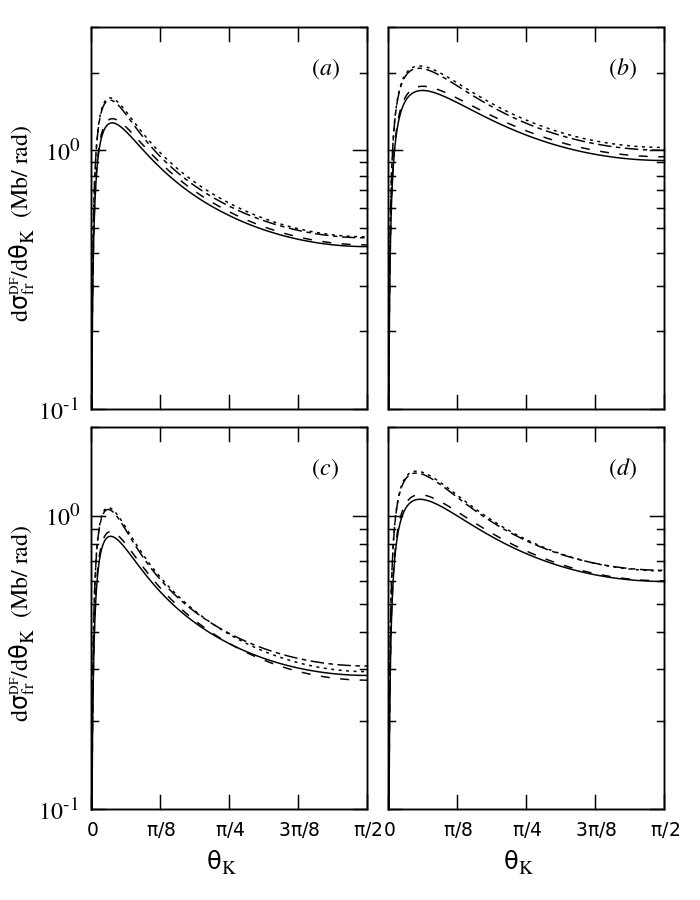}} 
\vspace{-0.55cm}
\caption{   The angular distribution, 
$ \frac{d \sigma^{\text{DF}}_{\text{fr}}}{ d \Theta_{\bm K} } $,  of the He$^+$ fragments produced in collisions with $1$ GeV/u (two upper panels) and $7$ GeV/u (two lower panels) U$^{92+}$ projectiles. The results shown in panels ($a$), ($c$) and 
($b$),($d$) were obtained by integrating over the energy intervals 
$ 0 $ eV $\leq E_K \leq 2$ eV and $ 0 $ eV $\leq E_K \leq 7$ eV, respectively. Dotted, dash-dotted, dash and solid curves represent results of the calculations performed in the nr-FBA, nr-SEA, r-FBA and r-SEA approximations, respectively.  }
\label{figure5}
\end{figure}

In figure \ref{figure5} we display  the angular distribution of the He$^+$ fragments represented by the cross section $ \frac{ d \sigma^{\text{DF}}_{\text{fr}}}{ d \Theta_{\bm K} } $. The cross section was obtained by integrating either over the energy interval $ 0 $ eV $\leq  E_K \leq 2 $ eV, in which the direct fragmentation mechanism is by far the dominant one, or over the broader range  
$ 0 $ eV $\leq  E_K \leq 7 $ eV, which covers essentially all energies which are possible in the fragmentation events caused by this mechanism   
(see figures \ref{figure1}-\ref{figure2}).  

Figure \ref{figure5} shows that the angular distribution of the He$^+$ fragments becomes significantly weaker dependent on the angle $\Theta_{\bm K}$ when their energy $E_K$ increases. This feature (which is already seen in figures 
\ref{figure3}--\ref{figure4}) can be understood by taking into account that, due to the relation $E_K = 1/R $, the He$^+$ fragments with larger energies emerge when the projectiles hit the He$_2$ dimers with smaller size $R$. In such collisions the DF mechanism is more efficient causing the He$_2$ breakup with significant probabilities even when the dimer is oriented at large angles with respect to the impact velocity. 

\begin{figure}[h!]
\vspace{0.0cm} 
\hspace{-0.75cm}
\subfigure{\includegraphics[width=9.5cm]{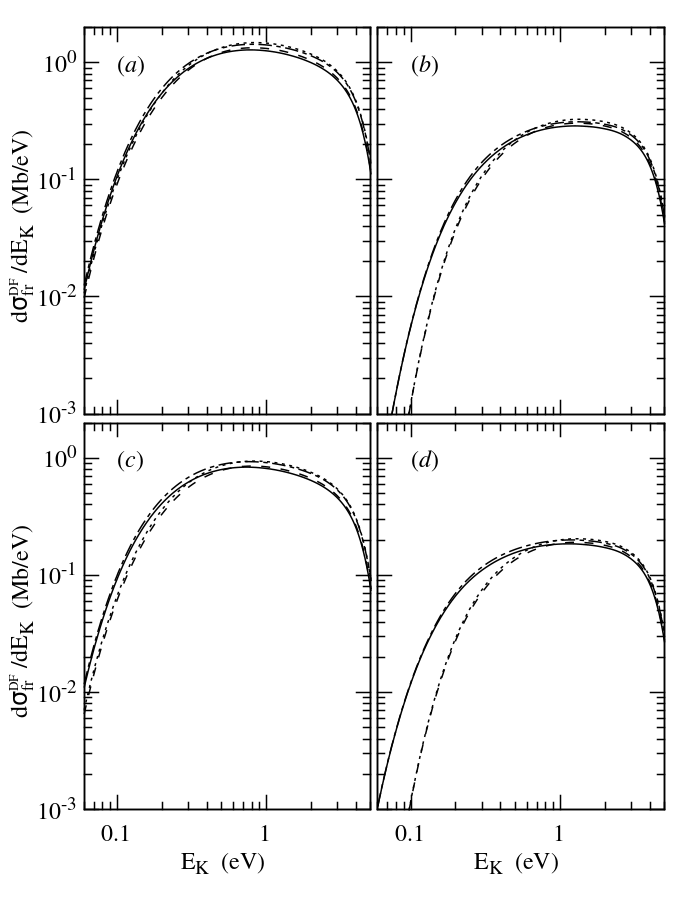}} 
\vspace{-0.55cm}
\caption{   The energy spectrum, 
$\frac{ d \sigma^{\text{DF}}_{\text{fr}} }{ dE_K }$, in the  fragmentation of the He$_2$ dimer in collisions with $1$ GeV/u 
(the upper panels) and $7$ GeV/u (the lower panels) U$^{92+}$ projectiles. 
The results shown in panels ($a$), ($c$) and ($b$), ($d$) were obtained by integrating over the angles $ 0^\circ \leq \Theta_{\bm K} \leq 180^\circ$ and $ 60^\circ \leq \Theta_{\bm K} \leq 120^\circ$, respectively. Dotted, dash-dotted, dash and solid curves represent results of the calculations performed in the nr-FBA, nr-SEA, r-FBA and r-SEA, respectively. }
\label{figure6}
\end{figure}
 
The energy spectrum, $d \sigma^{\text{DF}}_{\text{fr}}/dE_K$,  of He$^+$ fragments produced by bombarding the He$_2$ dimer by 
$1$ and $7$ GeV/u U$^{92+}$ projectiles is displayed in figure \ref{figure6}.  
The spectrum shown in the left panels of this figure was obtained by integrating over all possible fragmentation angles,  
$0^\circ \leq \Theta_{\bm K} \leq 180^\circ$, whereas that given in the right panels includes only the fragments 
moving at large angles, $60^\circ \leq \Theta_{\bm K} \leq 120^\circ$, with respect to the projectile velocity. 

The results shown in 
figure \ref{figure6} confirm the correspondences between the relativistic and higher-order effects and the energy $E_K$ and angle $\Theta_{\bm K}$ of the fragments which were discussed above. In particular, we again observe  
that at a given collision velocity 
the relativistic effects rise when the energy $E_K$ decreases, becoming especially strong in collisions where the He$^+$ fragments possess quite small energies $E_K$ and move at large angles with respect to the projectile velocity. 

\vspace{0.25cm} 

\begin{figure}[h!]
\vspace{0.0cm} 
\hspace{-0.75cm}
\subfigure{\includegraphics[width=9.2cm]{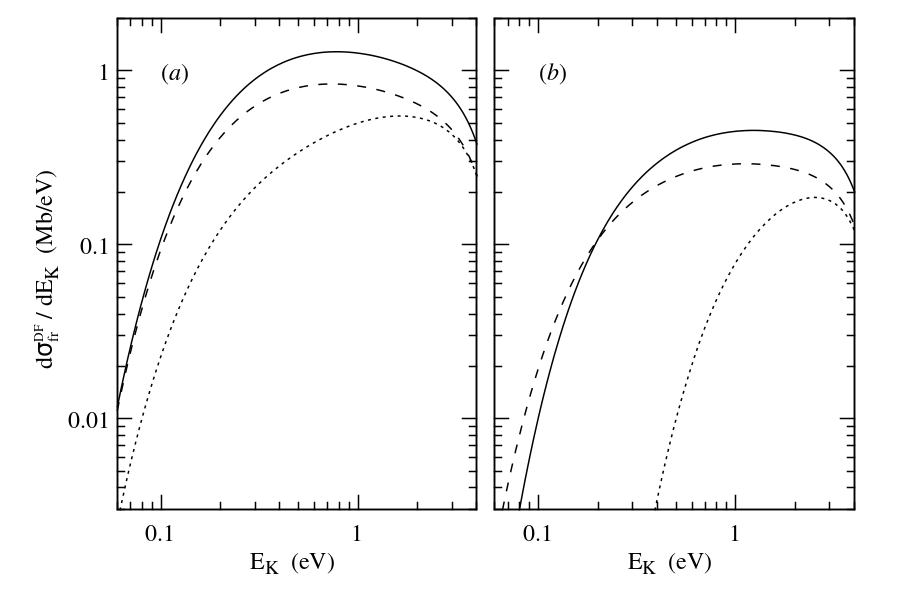}} 
\vspace{-0.55cm}
\caption{   The energy spectrum, 
$\frac{ d \sigma^{\text{DF}}_{\text{fr}} }{ dE_K }$, for the He$_2$ breakup into He$^+$ via the DF mechanism calculated within the r-SEA. 
Results for collisions with $1$ GeV/u U$^{92+}$, 
$7$ GeV/u U$^{92+}$ and $11.37$ MeV/u S$^{14+}$ are shown by solid, dashed and dotted curves, respectively. The results shown in ($a$) and ($b$) were obtained by integrating over $ 0^\circ \leq \Theta_{\bm K} \leq 180^\circ $ and $ 45^\circ \leq \Theta_{\bm K} \leq 135^\circ $, respectively. }
\label{figure7}
\end{figure}

In figure \ref{figure7} we compare the energy spectra of the He$^+$ ions produced in collisions with $1$ and $7$ GeV/u U$^{92+}$ and $11.37$ MeV/u S$^{14+}$ ($v = 21.2$ a.u., $\gamma = 1.01 $). It is seen that the shape of the spectrum profoundly varies with an increase in the impact energy. In particular, the maximum of the energy distribution is shifted from $ E_K \approx 1.7$ eV in collisions with $11.37$ MeV/u S$^{14+}$ to $E_K \approx 0.7$ eV in collisions with $7$ GeV/u U$^{92+}$ where the low-energy part of the spectrum is strongly enhanced 
(see figure \ref{figure7}a) and this enhancement becomes even much more pronounced if the fragmentation events with large angles $ \Theta_{\bm K} $ are selected (figure \ref{figure7}b). 

In collisions with $11.37$ MeV/u S$^{14+}$ and 
$7$ GeV/u U$^{92+}$ the effective strength of the projectile field, given by the ratio $Z_p/v$, is essentially the same ($ 0.68 $ and $ 0.66 $, respectively). Therefore, the strong enhancement of the lower-energy part of the fragmentation spectrum, which reflects the corresponding strong increase in the breakup of dimers with very large instantaneous size $R$,  is caused by 
a much higher impact energy. One should note, however, that since He$_2$ is a very light target, 
a large increase in the collision velocity has a more profound overall effect on the spectrum shape and the total number of events than an increase in the Lorentz factor $ \gamma $.   

\subsection{ The energy spectrum shape versus the He$_2$ binding energy }

\begin{figure}[h!]
\vspace{0.0cm}
\centering 
\subfigure{\includegraphics[width=7.cm]{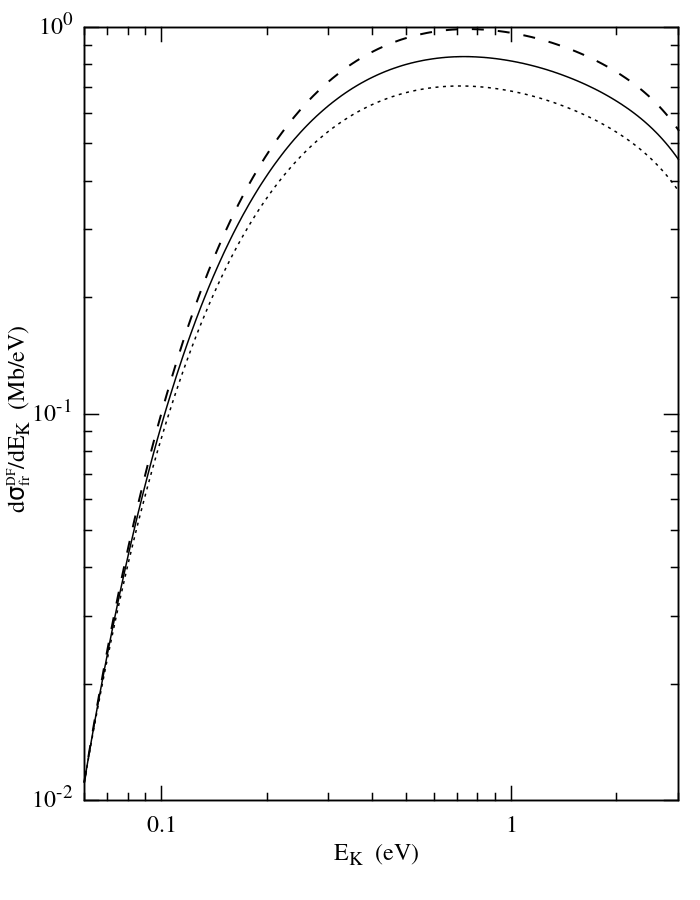}} 
\vspace{-0.55cm}
\caption{ The cross section $\frac{ d \sigma^{\text{DF}}_{\text{fr}} }{ dE_K }$, versus the binding energy $I_b$ of the He$_2$ in collisions with $7$ GeV/u U$^{92+}$ projectiles. Solid, dashed and dotted curves correspond to $I_b = 139 $ neV, $I_b = 148 $ neV and $I_b = 130$ neV, respectively. The results for $I_b = 148 $ neV and $I_b = 130$ neV were normalized to those for $I_b = 139 $ neV at $E_K = 60$ meV. }
\label{figure8}
\end{figure}

The values for the binding energy $I_b$ of the He$_2$ dimer, reported in the literature, 
vary between $I_b = 44.8$ neV \cite{44.8} and 
$ I_b = 161.7$ neV \cite{161.7}. The distribution of the probability density $ \rho(R) = \vert \Psi_i(R) \vert^2$ in the ground state of the dimer depends on the value of $I_b$ and a variation 
$ \Delta I_b $ in the binding energy would be  reflected by the corresponding variation $\Delta \rho(R)$ of the probability density $ \rho(R) $. However, if $ \Delta I_b $ is relatively small, 
$ \Delta I_b \ll I_b$, a large range of $R$ has to be spanned in order for the changes in the shape of $ \rho(R)$ to become noticeable.   
    
In this respect, as it was already emphasized in \cite{He2-rel-HCI-lett}, the fragmentation by ultra-fast projectiles can be of especial interest.   
Indeed, since the adiabatic collision radius 
$ R_a $ increases with the impact energy as $ \sim v \gamma $, they possess a very large effective interaction range that is a great advantage in probing the structure of such enormous objects like the He$_2$ dimer. 
 
In figure \ref{figure8} we display the energy spectrum 
$\frac{ d \sigma^{\text{DF}}_{\text{fr}} }{ dE_K }$ for the He$_2$ fragmentation by $7$ GeV/u U$^{92+}$. In the range of $E_K$,   
shown in the figure, the direct fragmentation is by far the dominant breakup channel and the cross section 
$\frac{ d \sigma^{\text{DF}}_{\text{fr}} }{ dE_K }$ represents the total energy spectrum. 
   
The spectrum was calculated for three values of the dimer binding energy $I_b$: 
$ 130 $ neV, $ 139 $ neV and $ 148 $ neV.  For a better visibility of the variation of the spectrum shape with $I_b$, the spectra for 
$ I_b = 130 $ neV  and $ I_b = 148 $ neV were normalized to the spectrum for $ I_b = 139 $ neV at an energy  $E_K = 60$ meV.      

In the energy interval,  $60$ meV  $ \leq E_K \leq  1$ eV, the ratios  
$\frac{ d \sigma^{\text{DF}}_{\text{fr}}(I_b=148) }{ dE_K }/  
\frac{ d \sigma^{\text{DF}}_{\text{fr}}(I_b=139) }{ dE_K }$, 
$\frac{ d \sigma^{\text{DF}}_{\text{fr}}(I_b=139) }{ dE_K }/  
\frac{ d \sigma^{\text{DF}}_{\text{fr}}(I_b=130) }{ dE_K }$ and 
$\frac{ d \sigma^{\text{DF}}_{\text{fr}}(I_b=148) }{ dE_K }/  
\frac{ d \sigma^{\text{DF}}_{\text{fr}}(I_b=130) }{ dE_K }$
vary by about $ 18 \%$, $ 20 \%$ and $ 42 \%$, respectively. The corresponding relative variations, 
$\Delta I_b/I_b$,  
of the binding energy  
are approximately $ 6.5 \%$, $ 7 \%$ and $ 14 \%$, respectively. This means that in the energy interval under consideration the shape of the energy spectrum 'magnifies' 
the variation of the binding energy by about three times. 

A more detailed analysis of the cross sections shows that the range of quite small $E_K$ ($E_K \lesssim 0.2$ eV) gives the main contribution to the variation of the cross section ratios. This is not very surprising since the small energy range,   
\, $60$ meV $ \leq E_K \lesssim 0.2$ eV, \, corresponds to a very large interval,  
$ 136 $ a.u. $ \lesssim  R  \lesssim 454 $ a.u., of the interatomic distances in the He$_2$ dimer, where the shape of its ground state becomes sensitive to even a very modest variation of the binding energy. 
The projectile is able to span in the collision so large interval of distances since at $ 7 $ GeV/u the adiabatic collision radius 
$R_a \simeq 10^3$ a.u. already significantly exceeds the size of the dimer.

\subsection{ The total fragmentation cross sections }
 
According to our r-SEA calculations, the contribution 
$ \sigma^{\text{DF}}_{\text{fr}} $ of the DF mechanism to the total cross section for the breakup of the He$_2$ dimer    
in collisions with $1$ and $7$ GeV/u U$^{92+}$ projectiles 
amounts to $3.65$ Mb and $2.39$ Mb, respectively. 

The relativistic effects in the direct fragmentation are most significant when the 
'instantaneous' transverse size $R_{\perp}$ of the He$_2$ dimer in the collision is very large. Consequently, they most profoundly influence the fragmentation events with small $E_K$ and large $\Theta_{\bm K}$. However, such events have relatively small probabilities and, as a result, the cross section  
$ \sigma^{\text{DF}}_{\text{fr}} $ 
is very weakly influenced by the relativistic effects. 
In particular, for collisions with $1$ and $7$ GeV/u U$^{92+}$ the difference between our results for $ \sigma^{\text{DF}}_{\text{fr}}$, obtained using relativistic and nonrelativistic approaches, is about merely $\simeq 1 \%$.  

The field of the projectile acting on the dimer is strongest when the impact parameters with respect to both atomic sites of the dimer are small, i.e. in collisions 
where the 'instantaneous' transverse size $R_{\perp}$ of 
the dimer is not large. Since such collisions give a more significant contribution to $\sigma^{\text{DF}}_{\text{fr}}$, than those with very large $R_{\perp}$, the influence of the higher-order effects in 
the interaction between the projectile and the dimer on the magnitude of $\sigma^{\text{DF}}_{\text{fr}}$ is substantially larger reaching about $14 \%$ for the fragmentation by $1$ and $7$ GeV/u U$^{92+}$.    

\vspace{0.25cm} 

In the present paper the He$_2$ fragmentation via the 
IE-ICD and DI-RET mechanisms was not calculated. Nevertheless, we can still obtain a rough estimate for the contribution to the total fragmentation cross section due to these two mechanisms. According to the experimental results of \cite{thesis_He2_S14+}, 
in the breakup of the He$_2$ dimer by 
$11.37$ MeV S$^{14+}$ the ratio of the summed contributions of the IE-ICD and DI-RET channels 
to the contribution of the DF channel 
is approximately equal to $1.6$. Taking into account 
that our calculation for the DF contribution in collisions with $11.37$ MeV/u S$^{14+}$ yields $1.8$ Mb, we obtain that the summed contributions of the IE-ICD and DI-RET to the fragmentation by $11.37$ MeV S$^{14+}$ is $ \approx 1.8 \times 1.6 = 2.9$ Mb.  

The cross sections for the fragmentation via the IE-ICD and DI-RET are proportional to the cross sections for simultaneous ionization-excitation and double ionization, respectively, of the helium atom. Both these cross section scale similarly with $Z_p$ and $v$. Therefore, by calculating them for collisions with $11.37$ MeV/u S$^{14+}$ and $1$ and $7$ GeV/u U$^{92+}$ projectiles and using our results for the DF cross sections, we have estimated that the summed contribution of the IE-ICD and DI-RET mechanisms to the fragmentation in collisions with $1$ and $7$ GeV/u U$^{92+}$ is roughly equal to 
$ 4.3 $ Mb and $ 3 $ Mb, respectively. 
 
\section{ Conclusions }

In conclusion, we have studied    
the fragmentation of the helium dimer into 
singly charged helium ions by relativistic highly charged projectiles in collisions with relatively low kinetic energy release 
$E_K \lesssim 3$--$4$ eV. Such breakup events occur solely due to the direct fragmentation mechanism in which 
the projectile  
ionized both dimer's atoms in a single collision. 
In this mechanism the two helium ions are produced within the so short time interval ($ \lesssim 10^{-16}$ s) that during 
the production the helium nuclei remain essentially at rest. Consequently, the energy $E_K$ of the ionic fragments, produced via the direct mechanism, is very simply ($E_K = 1/R $) related to the size $R$ of the He$_2$ dimer at 
the collision instant. 
 
We have investigated in detail the energy and angular spectra of the He$^+$ ions produced in collisions with $1$ and $7$ GeV/u U$^{92+}$ projectiles. Our main findings can be summarized as follows.  

In collisions with $\gamma \gg 1$ the fragmentation events, in which the He$^+$ ions move with low kinetic energies 
($E_K \lesssim 0.1$ eV) under large angles ($ 20^\circ \lesssim \Theta_{\bm K} \lesssim 160^\circ $) with respect to the projectile velocity, are strongly affected by the relativistic effects.  

The shape of the energy spectrum of the He$^+$ ions is quite sensitive to the binding energy of the He$_2$ dimer which can be exploited for its precise determination. Here the relativistic effects also play the important role since, by significantly enhancing the lower-energy part of the spectrum, they enable one to span a substantially broader range of the inter-atomic distances $R$ in the dimer, effectively increasing the sensitivity of the spectrum shape to the variation of the dimer binding energy.  

The relativistic effects, having a strong impact on the fragments with low $E_K$ and large $\Theta_{\bm K}$ and making the energy spectrum shape more sensitive to the variation in the dimer binding energy, influence nevertheless very weakly the total amount of the fragmentation events caused by the direct mechanism 
because they affect only their minor part. 
In this respect, a large increase in the collision velocity has a much stronger overall effect on the spectrum shape and the total number of events. 

In contrast to relativistic effects, the role of the higher-order effects in the projectile-dimer interaction rises with increasing energy of the He$^+$ fragments becoming substantial at $E_K \gtrsim 1$ eV. Since such events give a more significant contribution to the fragmentation than those with very small $E_K$, the influence of the higher-order effects is substantially larger reaching about $14 \%$ for the total cross section in the direct fragmentation by $1$ and $7$ GeV/u U$^{92+}$.    

According to our calculations the contribution of the direct mechanism to the total cross section for the He$_2$ fragmentation by $1$ and $7$ GeV/u U$^{92+}$ amounts to $3.65$ Mb and $2.39$ Mb, respectively. A rough estimate for the total fragmentation cross section in these collisions, which takes into account all the fragmentation mechanisms, suggests that it 
is about two times larger than the above values.

\begin{acknowledgements}
BN, SFZ and XM gratefully acknowledge 
the support from the ‘National Key 
Research and Development Program of China’ (Grant No. 2017YFA0402300) and the CAS President’s Fellowship Initiative. 
Our numerical results were obtained using the facilities of the Supercomputer Center HIRFL at the Institute of Modern Physics (Chinese Academy of Sciences).
\end{acknowledgements}


\begin{thebibliography}{33}

\bibitem{He2_alpha-particle} 
J. Titze, M. S. Sch\"offler, H.-K. Kim, et al, 
Phys. Rev. Lett. {\bf 106}, 033201 (2011). 

\bibitem{He2_S14+}
H.-K. Kim, H. Gassert, J. N. Titze, et al, 
Phys. Rev. {\bf A 89} 022704 (2014); 

\bibitem{thesis_He2_S14+}
H.-K. Kim, PhD Thesis, (Frankfurt University, 2014)  (unpublished).  

\bibitem{He-Li-rel-HCI} A. Jacob, C. M\"uller and A: B. Voitkiv, Phys. Rev. {\bf  A 103}, 042804 (2021). 

\bibitem{He2-rel-HCI-lett} B. Najjari, Z. Wang 
and A. B. Voitkiv, 
Phys. Rev. Lett. {\bf 127}, 203401 (2021). 

\bibitem{e-l} R. E. Grisenti, W. Sch\"ollkopf, J. P. Toennies, G. C. Hegerfeldt, T. K\"ohler, and M. Stoll,   
Phys Rev Lett {\bf 85}, 2284 (2000). 

\bibitem{14A} B. A. Friedrich, 
Physics {\bf 6}, 42 (2013).

\bibitem{Cas-Pol} H. B. G. Casimir and D. Polder, 
Phys. Rev. {\bf 73}, 360 (1948).

\bibitem{retard} F. Luo, G. Kim, G. C. McBane, C. F. Giese, and W. R.
Gentry, J. Chem. Phys. {\bf 98}, 9687 (1993); 
M. Jeziorska, W. Cencek, K. Patkowski, B. Jeziorski, and
K. Szalewicz, J. Chem. Phys. {\bf 127}, 124303 (2007).

%

\bibitem{He2_photon} 
T. Havermeier, T. Jahnke, K. Kreidi, et al, 
Phys. Rev. Lett. {\bf 104}, 153401 (2010). 

\bibitem{He2_photon_icd} T. Havermeier, 
T. Jahnke, K. Kreidi, et al,  
Phys. Rev. Lett. {\bf 104}, 133401 (2010). 

\bibitem{He2_photon_icd-calc} 
N. Sisourat , N. V. Kryzhevoi1 , P. Kolorenc, 
S. Scheit, T. Jahnke and L. S. Cederbaum, 
Nature Physics {\bf 6}, 508 (2010).  

\bibitem{dimer-binding-exp} S. Zeller, M. Kunitski, J. Voigtsberger, et al, www.pnas.org/cgi/doi/10.1073/pnas.1610688113 (2016). 


\bibitem{el-cpt} 
For example, in collisions of 
$1$ GeV/u U$^{92+}$ with He  
the radiative electron capture cross section   
$ \sigma_{rec} \simeq 10^{-23}$ cm$^2$ 
(see e.g. results for radiative recombination on p. 272 of \cite{eic}) 
is about $8$ orders of magnitude smaller 
than the cross section for single ionization of He  
$ \sigma_{i} \approx 10^{-15}$ cm$^2$ 
(see e.g. \cite{He-ioniz}).  
The process of nonradiative electron capture is characterized by even much weaker cross sections.     

\bibitem{eic}  J. Eichler and W. E. Meyerhof, 
{\it Relativistic Atomic Collisions}
(Academic Press, San Diego, 1995).  

\bibitem{we-1998} A. B. Voitkiv and A. V. Koval, 
J. Phys. {\bf B 31}, 499 (1998).   

\bibitem{He-ioniz} A. B. Voitkiv, B. Najjari, R. Moshammer, and J. Ullrich, Phys. Rev. {\bf A 65}, 032707 (2002).  

\bibitem{we-21-reflect-appr} 
B. Najjari and A. B. Voitkiv,   
Phys. Rev. {\bf  A 104} 033104 (2021). 

%
\bibitem{i-a-a} J. Matthew and Y. Komninos, 
Surf. Sci. {\bf 53}, 716 (1975);   

%
\bibitem{icd} 
L. S. Cederbaum, J. Zobeley, and F. Tarantelli, Phys. Rev.
Lett. {\bf 79}, 4778 (1997); 
R. Santra, J. Zobeley, L. S. Cederbaum, and N. Moiseyev, Phys. Rev. Lett. {\bf 85}, 4490 (2000).   


\bibitem{Sulf} This mechanism was initially considered in \cite{He2_S14+} for collisions of $11.37$ MeV/u 
S$^{14+}$ with He2 dimers having the 'instantaneous' internuclear distance $R$ in the range $ 5 $ a.u. $ \leq R \leq 10 $ a.u.   

\bibitem{f1} For the He$_2$ fragmentation 
by photo absorption a similar mechanism was explored experimentally and theoretically in  
\cite{He2_photon} and \cite{we-21-reflect-appr}, respectively. 


\bibitem{new-approach} In \cite{He2-rel-HCI-lett} the probabilities $ P_{A^+}(\bm b)$ and $ P_{B^+}(\bm b')$ were defined as $ P_{A^+}(\bm b)  =  2 \, w(\bm b) $ and $ P_{B^+}(\bm b') = 2 \, w(\bm b') $. The inclusion of the factors $ (1 - w(\bm b))$ and  
$ (1 - w(\bm b')) $ somewhat reduces  
the cross section values compared to those reported in \cite{He2-rel-HCI-lett}.  



\bibitem{param} F. Martin and A. Salin, 
Phys. Rev. {\bf A 55}, 2004 (1997).

\bibitem{A-S} M. Abramowitz and I. Stegun, 
{\it Handbook of Mathematical Functions} 
(New York: Dover, 1964). 

\bibitem{param-C} For instance, in the interval $ 50 $ a.u. 
$ \lesssim R_\perp \lesssim 500 $ a.u. the value of $C$ varies by just about $16 \%$ (between $0.56$ and $0.65$ if $R_\perp$, $R_a$, $Z_p$ and $v$ are given in a.u. while the cross section is in Mb).     


\bibitem{rhci} R. Moshammer et al, 
Phys. Rev. Lett. {\bf 79}, 3621 (1997).  

\bibitem{r-sea} A. B. Voitkiv and B. Najjari, 
J. Phys. {\bf B 37}, 4831 (2004).

\bibitem{139.2} M. Przybytek et al, 
Phys. Rev. Lett. {\bf 104}, 183003 (2010).

\bibitem{nr-sea} P. D. Fainstein and R. D. Rivarola, J. Phys. {\bf B 20}, 1285 (1987). 

\bibitem{we-jpb-2005} A. B. Voitkiv, B. Najjari and J. Ullrich, J. Phys. {\bf B 38}, L107 (2005).  
 
\bibitem{44.8} R. Feltgen, H. Kirst, K. A. K\"ohler et al, 
J. Chem. Phys. {\bf 76}(5), 2360 (1982).

\bibitem{161.7} A. R. Janzen, R.A. Aziz, 
J. Chem. Phys. {\bf 107}(3), 914 (1997).


\end{thebibliography}
\end{document}